# The life cycle of scientific principles
# – a template for characterizing physical principles

Radin Dardashti, Enno Fischer, Robert Harlander

**Abstract**

Scientific principles can undergo various developments. While philosophers of science have acknowledged that such changes occur, there is no systematic account of the development of scientific principles. Here we propose a template for analyzing the development of scientific principles called the 'life cycle' of principles. It includes a series of processes that principles can go through: prehistory, elevation, formalization, generalization, and challenge. The life cycle, we argue, is a useful heuristic for the analysis of the development of scientific principles. We illustrate this by discussing examples from foundational physics including Lorentz invariance, Mach's principle, the naturalness principle, and the perfect cosmological principle. We also explore two applications of the template. First, we propose that the template can be employed to diagnose the quality of scientific principles. Second, we discuss the ramifications of the life cycle's processes for the empirical testability of principles.

**Keywords:** Scientific principles, Lorentz invariance, Mach's principle, naturalness, testability

# 1    Introduction

Scientific principles guide our searches for new theories and models. They can act as an important motivation for generating new theoretical proposals and they can impose constraints on future theories. So, scientific principles have an impact on how theories change and how they develop. But principles themselves, and their significance for physics, also change and develop. A particularly prominent discussion of this process can be found in Friedman's 'Dynamics of Reason' (2001), where he addresses the idea of a relativized a priori. A concrete case study has been provided by Massimi (2005), for example, who analyzes the various stages that the Pauli exclusion principle has gone through.

However, a systematic study of the development that scientific principles go through has not yet been provided. In this paper, we will attempt a first step in this direction. More specifically, we will present a template for analyzing such development named the "life cycle of scientific principles". It lists five processes that principles can go through: prehistory, elevation, formalization, generalization, and challenge.

After introducing the life cycle and explaining it on the basis of a few examples, we consider two applications for this framework. First, we will argue that the five processes it contains can be understood as *diagnostic criteria* for the quality of a principle. Like medical symptoms indicate information about the health status of a person, the processes of the life cycle provide useful information about the quality of a principle. Obviously, like the full determination of a medical condition should not be based on external symptoms alone, the proper judgement of the quality of a scientific principle still requires taking into account specific features of the principle under consideration and considerable practitioner's skills. However, we believe that the life cycle of scientific principles provides a useful diagnostic tool.

Second, we will look at the *testability* of scientific principles. For this purpose, we introduce the *theory space* associated with a certain principle as the set of theories that satisfy it. By employing a confirmation-theoretic framework we will show that the processes described by the life cycle have important ramifications for the testability. For example, formalizing a scientific principle may help define more clearly the boundaries of the associated theory space. This, in turn, helps identify the kinds of empirical evidence that would speak for or against that principle, or in other words, it increases its testability.

It should be noted that the concept of scientific principle is notoriously vague, and that it is employed by physicists to refer to a wide variety of items. Here we will not try to provide a unified and clear-cut definition of a scientific principle because such a definition may unduly limit the scope of our discussion. Instead, we believe that our systematic study of the life-cycle template may help us to understand the character and development of scientific principles.

The outline of the paper is as follows. In section 2, we will present the idea of the life cycle of scientific principles. We will address the individual processes in turn and will provide a few clarificatory remarks regarding the purpose and scope of the life-cycle template. In section 3, we will look at a number of physical principles and address in what sense and to what degree the development of these principles has undergone the processes of the life cycle. In section 4, we will address the life cycle as a diagnostic tool. In section 5, we will relate the life cycle to the testability of scientific principles.

## 2    The life cycle of principles

Principles are not static features of scientific reasoning. They undergo certain developments and change their properties and role in theory development. Principles may lead to empirically adequate theories at one time while being of hindrance to theory development at another time. This context-dependent and developing character is an important aspect of scientific principles, which is recognized widely in the history of physics literature (Massimi 2005; Darrigol 2021). But it has not been recognized to the same degree in the philosophy of science literature. The latter has been mostly concerned with attempts at providing a taxonomy of principles (e.g., Crowther 2021), or the classification of principles according to a certain status (e.g., constitutive principles in Friedman 2001). We argue (see Sect. 5) that these approaches are enriched once one takes into account that their specific philosophical assessment may be linked to the state that the principle currently assumes in the life cycle.

In this section, we introduce certain processes that a scientific principle can undergo from its conception to its possible demise. We therefore call this the life cycle of scientific principles. We should address some potential misunderstandings right away. First, the life cycle is not to be understood as a deterministic description of the development of scientific principles. Principles may undergo the processes described here to a higher or lower degree. In fact, we believe that interesting differences between specific principles can be seen by looking at how they differ in terms of the processes that they have gone through. Moreover, with regard to any specific process, there may arise considerable disagreement among scientific practitioners whether a principle has gone through the process or will ever go through it. Second, the various processes are not necessarily disjoint. For example, the process of formalization may coincide with the act of elevation, as in the example of the Heisenberg uncertainty principle. Third, we also do not claim that the list of processes defined here is complete, nor that every principle has to undergo all processes. In particular, the term life cycle does not imply that every principle will "die" at some point. The processes that we identify represent crucial steps in the development of a scientific principle, and they can be linked to possible epistemic advantages of a principle (see Sect. 4 and Sect. 5). So, aside from the obvious descriptive component, there is also a normative goal underlying the life cycle, as we aim to evaluate principles based on the processes it contains.

## 2.1  Prehistory

Scientific principles do not emerge from thin air. The potential role and features of a principle are already recognized in its prehistory. It is also in the prehistory of a principle where its initial justification can be found and thus within which the initial commitment of the scientists can be understood. Darrigol, for example, emphasizes the importance of such prehistoric developments in the context of relativity principles. He argues that the prehistory of such principles "directly inspired some of Poincaré's reflections toward a theory of relativity, including the name he gave to the relativity principle; it indirectly informed Einstein's reflections through some of Poincaré's; it plausibly contributed to the genesis of the equivalence principle; and it belonged to a rising physics of principles in which Einstein inscribed his own efforts" (2021, 51).

As a look at a broader sample of examples shows, there does not seem to be a common kind of prehistory associated with the emergence of a principle. Different features of a principle and the potential role of the principle can emerge in the prehistory through several routes. Here are some examples:

**A principle may be a feature of an already existing theory:** Scientists may for instance recognize a principle as a crucial feature of an already existing successful theory. An example is Lorentz invariance, the principle which implies that the laws of physics are the same in all inertial frames. Another example is the gauge principle that states a connection between local gauge symmetries and vector fields. That the related gauge invariances are present in Maxwell's electrodynamics was known already in the early 20th century but not given much attention until Hermann Weyl's work on unified field theories (O'Raifeartaigh and Straumann 2000; Jackson and Okun 2001; Berghofer et al. 2023, 15ff).

**Empirical support:** There may be new or old experimental results that support the principle. Think of the experimental results of spectroscopy that lead up to the Pauli exclusion principle (Massimi 2005) or the Michelson-Morley experiment that can be understood as supporting the principle of constancy of the speed of light.

**Meta-inductive support:** Certain principles gain in popularity only over time. The naturalness principle in high-energy physics is often motivated in this way. Meta-inductive support can accrue during a principle's prehistory, that is at a stage when it is not recognized as a principle. But the reconstruction of past successful applications is then often taken as supporting the elevation of the principle in hindsight (for the case of the naturalness principle see (Carretero Sahuquillo 2019; Borrelli and Castellani 2019)).

**Support through explanatory coherence:** Recognizing how a certain feature of a theory would increase the overall explanatory coherence of a theory does often provide support for the elevation of a principle. The elevation usually would go hand in hand with a simultaneous reduction of the number of brute facts of the theory. For example, the introduction of the equivalence principle in General Relativity explains the equivalence of inertial and gravitational mass, which remains an unexplained fact in Newtonian Gravity (see (Janssen 2002)).

**Metaphysical commitments:** There may be metaphysical presuppositions that speak in favor of a principle. Consider the perfect cosmological principle, which requires the universe not only to be homogeneous and isotropic as in the cosmological principle but also static (Hoyle 1948). After the observation of the Hubble red-shift, which supports an expanding universe, a static universe seemed less likely. Yet, Hoyle (1948) required the perfect cosmological principle in order to establish how the universe should be. So it was despite rather than due to the empirical evidence that they required the perfect cosmological principle to hold. We will discuss this example in more detail in Sect. 3.4.

**Thought experiments:** Principles may also find initial support in the form of thought experiments. This was the case with Galileo's thought experiment about experiments on a ship in support of the relativity principle or Einstein's elevator thought experiment in support of the equivalence principle.

**Forward-looking justification:** Sometimes principles gain support through their relation to promising theories. The naturalness principle of high-energy physics, for example, was motivated through its relation of potential coherence with promising yet unconfirmed theories of Beyond the Standard Model (BSM) physics (Fischer 2023).

The various routes may provide support in various ways and in different strengths. Moreover, no single reason is usually used in isolation for an elevation. It is often the combination of several reasons that lead to the elevation of a certain feature to a principle of a theory. For instance, as we saw above, it was not only the explanatory coherence of the equivalence principle that led to its elevation, but also the plausibility it gained through thought experiments.

## 2.2    Elevation

Elevation describes the process in which a claim gains the status of a scientific principle. In some instances, elevation can be understood as a declarative act or a baptism. Lorentz invariance, for example, was a well-known feature of electrodynamics; it was a recognized fact with empirical support before it gained the status of a principle. It was elevated to the status of a principle only later through Einstein's employing it to "*define* a fundamentally new notion of simultaneity" (Friedman 2001, 88, emph. original).[1] In other instances, elevation describes a process extended in time, a process that is not necessarily initiated or concluded by the actions of a single scientist. Massimi (2005), for example, argues that Pauli exclusion accrued the status of a principle through being incorporated into the new quantum mechanics.

---

[1] More precisely, Friedman argues that "[w]hereas Lorentz and Fitzgerald take an essentially classical background structure for space, time, and motion to be already sufficiently well-defined and only subsequently locate the new empirical discovery in question as a peculiar (but additional) empirical fact formulated against the background of this classical structure, Einstein calls the whole classical structure into question and uses the very same empirical discovery empirically to define a new fundamental framework for space, time, and motion entirely independently of the classical background. It is in precisely this way [...] that Einstein has 'elevated' an empirical law to the status [...] of a coordinating or constitutive principle" (Friedman 2001, 88). We discuss the example of Lorentz invariance in more detail in section 3.1.

What does the process of elevation amount to? That is: what exactly happens to a scientific principle if it is being elevated? It is plausible to assume that by being elevated, a principle gains a certain epistemic status: it is to a certain degree shielded against challenges.

At least in some salient examples, elevation may be understood as a *hypothetical prioritization*. We describe a prioritization as hypothetical if an assumption is endowed with its privileged status while retaining its conjectural status. 'Hypothetical' may here be understood in two ways: as an evaluation of the principle in hindsight when the principle turned out to be problematic. Alternatively, 'hypothetical' can be understood as a term that describes the principle's acknowledged preliminary character in the context of current and ongoing research. Sometimes scientists explicitly flag the conjectural status of assumptions that are being elevated to principles. Einstein, for example, when introducing the principle of relativity, describes it as a "conjecture" ("Vermutung") that is to be elevated to the status of a "prerequisite" ("Voraussetzung") of his inquiry (Einstein 1905, 891). Likewise, Gerard 't Hooft (1980) makes the conjectural yet inquiry-dominating status of the naturalness principle quite explicit. According to 't Hooft, the naturalness principle captures the idea that it is "unlikely that the microscopic equations contain various free parameters that are carefully adjusted by Nature to give cancelling effects such that the macroscopic systems have some special properties." He identifies this as a "philosophy" to be applied to unified gauge theories. The conjectural yet inquiry-dominating or dogmatic character of the naturalness principle is expressed even more clearly when 't Hooft declares his version of the naturalness principle as prohibiting small parameters that are not protected by a symmetry: "*We now conjecture that the following dogma should be followed*: – at any energy scale μ, a physical parameter or set of physical parameters $\alpha_i(\mu)$ is allowed to be very small only if the replacement $\alpha_i(\mu) = 0$ would increase the symmetry of the system" ('t Hooft 1980, emphasis added).

There is a question of what causes the process of elevation. The answer to this question can often be found in the prehistory of the principle, as discussed above.

One should also note that the elevation in terms of a declarative act of a scientist does not need to entail that the principle is commonly accepted among all or even a majority of the members of a scientific community. The hypothetical character of the prioritization may vary among scientists depending partly on the various commitments they adhered to during the prehistory of the principle.

## 2.3    Formalization

Formalization describes the process in which a principle is put into formal terms. Here we illustrate the concept of formalization by highlighting that a principle can be formalized to different degrees. We begin (i) by discussing principles that lack a formalization. We contrast this with (ii) principles that have been cast into a "principle framework", which we take to be a particularly strong kind of formalization. Between the absence of a formalization and the principle framework there are (iii) various intermediate ways of formalization. At this point, it is important to clearly distinguish between "formulation" and "formalization" of a principle. A formulation does not necessarily involve mathematics. By contrast, formalization always involves some degree of mathematics.

(i) Some principles may initially not be formalized at all. Take the concept of relativity as introduced by Galileo Galilei in his *Dialogue Concerning the Two Chief World Systems.* Galileo does not refer to the idea of relativity as a principle. Instead, he considers a thought experiment, where someone is experimenting within a moving ship. He states the idea as follows:

[S]o long as the motion is uniform and not fluctuating this way and that, you will discover not the least change in all the effects named, nor could you tell from any of them whether the ship was moving or standing still. (Galilei 1632/1914, 187)

Galileo used this idea to explain why terrestrial experiments can't be used to show the movement of the earth ("all experiments practicable upon Earth are insufficient measures for proving its mobility"). So it plays an explanatory role even before it has been formalized in terms of, e.g., invariance under the Galilean transformations.

Another example is the early development of Pauli's exclusion principle. The Pauli exclusion principle began as a purely heuristic prescription in the old quantum theory for the occupation of energy levels in atoms by electrons. It was not yet formalized or elevated to the status of a principle but was considered a rule (*Ausschließungsregel*) (Massimi 2005).[2]

(ii) A particularly strong way of formalizing a principle is what we denote as a "principle framework". A principle framework constitutes a mathematical toolbox which ensures that a theory satisfies the principle if it is constructed completely within this framework.

Unlike Pauli's initial introduction of his exclusion rule, the new quantum theory allows one to express the principle more fundamentally in terms of the symmetry properties of wave functions. The complete theoretical framework for the principle became available in quantum field theory and its concept of creation and annihilation operators for particles. In this case, the anticommuting property of the fermionic operators embeds the Pauli principle into the theory. This was even more ingrained into the structure of the theory as part of axiomatic approaches to quantum field theory, where the Pauli principle is a consequence of the axioms of quantum field theory (see Bain 2016).

Relativity constitutes another prime example for such a case. All elements of the theory (space-time, momenta, field strengths, etc.) are associated with mathematical objects (tensors) which are defined by their transformation properties under the change of the reference frame. Well-defined rules for operations with these tensors ensure that any theory constructed within this mathematical framework is guaranteed to meet the principle of relativity. In the case of relativity, the formalism had already been developed in the mathematical literature by Ricci-Curbastro and Levi-Civita at the end of the 19th century. In other cases, the quest for such a formalism established new mathematical concepts. A prominent example here is Newton's infinitesimal calculus which he developed in parallel to his principles in the *Principia*.

---

[2] Note that a principle can be elevated even if it has not yet been formalized. Consider the elevation of Mach's principle by Einstein within the context of General Relativity. Einstein states the principle as the requirement that "the G-field is completely determined by the masses of the bodies" (Einstein 1918). The "G-Field" is what we now refer to as the Einstein tensor, and it encodes information about the geometry of space and time. Thus, Mach's principle acts as a non-formalized principle which, if enforced, would, for example, rule out the vacuum solutions of the Einstein field equation as possible physical solutions.

Sometimes it can be the framework itself that gives rise to the principle in the first place. This was the case for the Heisenberg uncertainty principle which is a consequence of the mathematical formalism of quantum mechanics.

(iii) The absence of a formalization and the principle framework can be seen as the extreme cases on a spectrum of various kinds of formalization. In what follows we consider some intermediate kinds of formalization. For example, principles can be put in terms of a mathematical condition that imposes constraints on theories. This is the case with Bohr's correspondence principle, which in its general formulation states that a quantum theory should recover classical physics in the limit of large quantum numbers. This can be translated into a numerical relation between the quantum mechanical frequency and the classical frequency for transitions at large quantum numbers.[3] As another example of a mathematically imposed constraint, consider the renormalizability of models in quantum field theory, which ensures predictivity at all energy scales. However, with the Wilsonian view of quantum field theories and the associated rise of effective field theories, which restrict predictivity to certain energy domains, renormalizability has lost its dominant role as a principle over the past few decades.[4]

In other cases of formalization, the mathematical formulation of the principle provides a well-defined criterion which allows one to test whether a particular theory obeys the principle or not. An example is the absence of anomalies in quantum theories. Anomalies arise when certain symmetries of the classical theory fail to be symmetries of the corresponding quantum theory. For gauge symmetries, this can lead to inconsistencies, which need to be tested for. The testability of a principle and its imposition on a theory may go hand-in-hand, in the sense that the principle selects a certain subset from a whole spectrum of theories. Absence of anomalies is a good example of this. While string theories could be formulated for an arbitrary number of space-time dimensions a priori, imposing the cancellation of anomalies restricts this number to 10 (or 26 for purely bosonic strings).

---

[3] It should be noted, though, that it has not always been clear what exactly the correspondence principle amounts to (Bokulich 2008; Bokulich and Bokulich 2020).

[4] See Rivat (2021) for the early history of Wilson's conception of effective field theories, Weinberg (1996) for a standard treatment of renormalization theory and Cao (2019) for a historical overview.

It should be noted that principles can undergo various transformations in terms of their formalization during their life cycle. On the one hand, one might trace the successful development of the relativistic nature of modern quantum field theories all the way back to Galileo's non-mathematized relativity principle. Its current realization, e.g. the axiomatic imposition in the formulation of quantum field theory, mirrors its foundational character and the degree one can trust its satisfaction in the accessible energy regimes. On the other hand, Mach's principle or the naturalness principle lack a similar formal development. The inherent vagueness of these principles has led to various, sometimes incompatible, formulations, where even once it has been explicated, it still leaves room for interpretation regarding whether developed theories satisfy them or not. We will return to and elaborate on these principles in Sect. 5.

## 2.4    Generalization and de-generalization

One may distinguish between two processes of generalization that a principle can undergo. On the one hand, one may consider one and the same principle, where one hypothetically extends the domain to which it is supposed to be applied to. Let us call this domain-generalization. On the other hand, one may consider one and the same domain of applicability, where however the principle itself has been generalized, in the sense that the initial principle now turns out to be a consequence of a more fundamental underlying principle. We call this principle-generalization. Of course, both kinds of generalizations can be and have often been realized simultaneously. That is, the principle-generalization of the principle also extended the domain to which it was supposed to be applied to. We will discuss both kinds of generalization now in more detail.

Domain-generalization is the process in which a scientific claim that is thought to hold true in one domain is extended to other domains. For instance, the application of the principle to one kind of entity may be extended to other kinds of entities. Most principles go through various steps of domain-generalization. Pauli's exclusion principle, for example, went through several steps that extended its domain of applicability. It was first introduced as a rule to describe spectroscopic phenomena, only at later stages was the principle acknowledged to hold true about phenomena in nuclear physics, atomic physics, condensed matter physics, and quantum chromodynamics. This development of the Pauli principle is not only concerned with the extension of the domain but went together with the recognition that the Pauli principle is a consequence of the more general spin-statistics theorem, which associates integer-spin particles with Bose statistics, and half-integer spin with Fermi statistics.

Scientific principles may also experience a process of domain-*de*generalization. In this process, a principle that is initially thought to hold true in a certain domain is found to hold true only in a smaller domain. For example, before the advent of relativity and quantum mechanics, the principles of Newtonian mechanics were thought to be principles that govern nature across all scales. With relativity and quantum mechanics being established, the laws of Newtonian mechanics still have the status of scientific principles, yet their domain of applicability is restricted to small velocities and macroscopic systems.

The process of generalization is a way in which principles fulfill their heuristic role. Suggesting the extension (or limitation) of a principle's domain of applicability is a way to generate new hypotheses. A common way to legitimize the naturalness principle, for example, is to point to various instances in which it previously held (and could have predicted the finding of new physics, see Carretero Sahuquillo 2019). From this it is hypothesized that theories should be natural in general. This, in turn, has led to the interpretation of the Higgs naturalness violation as an indicator for BSM physics in the TeV regime. Albeit not necessary, formalization may be a good basis for generalization. It may highlight the abstract and structural features of a principle that may be ready to be transferred to other domains of applicability. Pauli's formalization of the gauge principle in electrodynamics as associated with the U(1) group was a prerequisite for the significant step enabling Yang and Mills to extend the concept towards the SU(2) group of isospin.

On the other hand, there are instances in which a generalization of a principle is achieved by giving up a particular way to formalize the principle. The aforementioned example of Bohr's correspondence principle illustrates this. Originally, it describes a numerical correspondence between the quantum mechanical frequency and the classical frequency for transitions at large quantum numbers. It has been extended in various ways, and among other things to a principle about the *conceptual* relation between classical mechanics and quantum mechanics (Bokulich and Bokulich 2020). Bohr's correspondence principle has also been generalized to the generalized correspondence principle, a principle requiring that "any acceptable new theory *L* should account for the success of its predecessor theory *S* by 'degenerating' into that theory under those conditions under which *S* has been well confirmed by tests" (Post 1971, 228). These generalizations appear to require that the original formalization suggested by Bohr be given up in favor of broader or conceptual correspondence. The generalized correspondence principle, since it is supposed to be applicable to any theory (possibly even non-physical ones), is a case of a domain generalization. But it is also a principle-generalization in the sense that it extends far beyond relations between the quantum and classical frequencies. It has been critically discussed to what extent this generality has to be strongly qualified and relativized to a certain domain in order to be applicable (Radder 1991).

## 2.5    Challenge

This final process considers the challenges that are being put forward against principles. There are various reasons to challenge a scientific principle and various methods to implement that challenge. First, one may simply challenge a scientific principle for experimental reasons. Empirical results can in certain circumstances put a principle under pressure, leading (sometimes very slowly) to initial challenges and possibly to its abandonment. This was the case with the perfect cosmological principle (Bondi and Gold 1948; Fred Hoyle 1948), which was at the heart of the steady state models of the universe. Empirical results, especially the discovery of the cosmic microwave background, led finally to the abandonment of the principle (Kragh 2014). A similar development can currently be observed with the naturalness principle and the discovery of the unnaturally small Higgs mass.

Second, one may wish to challenge a principle due to proposed theoretical ideas and theories that may violate the principle in one way or another. The idea of a minimal length scale (e.g. Amelino-Camelia 2001) and certain features in current theories of quantum gravity often suggest the possibility of a violation of Lorentz invariance at the Planck scale (string theory, loop quantum gravity) or explicitly violate it (e.g., in non-commutative field theories) with possible observable consequences at accessible energy scales (see Mattingly (2005); Liberati (2013)).

There is no claim regarding the completeness of this list of reasons nor are these mutually exclusive in any sense. It is often likely that these challenges appear together. Moreover, there are various ways to implement these challenges. First and foremost, one may simply develop models which do *not* exhibit the principle and try to confirm them experimentally (as in the example of theories with a minimal length scale). Alternatively, one may challenge a principle with model-independent approaches. See Tasson (2014) for a review in the case of model-independent tests of Lorentz invariance.

## 2.6    Clarifications

It will be worth emphasizing once more that the list of processes is not to be understood as describing a deterministic development of principles. Apart from the prehistory, the processes may take place in varying order. The list of processes is also not exhaustive and there may be considerable overlap between them. So, what is the purpose of distinguishing these processes? We propose the life cycle as a heuristic template for analyzing the development of individual principles. Given a particular principle, can we identify the processes of prehistory and elevation? To what degree has the principle gone through the processes of formalization, generalization, challenge? Is there agreement among physicists about this? If the principle has not gone through some of these processes, why not? So, one important role of this template is to provide an understanding of the different stages of an individual principle's development.

This may help to classify individual scientific principles and gain a clearer view of the differences between scientific statements that have been identified as principles. It should be emphasized, though, that comparisons between potentially competing principles are not the concern of the paper.

# 3    Examples

## 3.1 Lorentz invariance

The principle of Lorentz invariance implies that the laws of physics are the same in all inertial frames. It underlies special relativity and plays an important role in large areas of contemporary fundamental physics. In what follows, we shall see that Lorentz invariance has gone through a series of processes as described by the life cycle in the foregoing section.

**Prehistory.** Features of Maxwell's electrodynamics that are associated with Lorentz invariance were known among physicists before Einstein's breakthrough contributions to special relativity (Brown 2005). The first Lorentz transformations were described by Larmor in 1900, who shows that Maxwell's equations for the free field are covariant under these transformations. Moreover, a concept of "local time" was employed by Lorentz, even though it had purely formal significance to him and was related to the problem of clock synchrony only by Poincaré. Furthermore, the experiments of Michelson and Morley strongly disfavored the existence of an ether which characterizes a specific frame of reference. Thus, Einstein's postulate of the principle of Lorentz invariance was indeed built on quite an elaborate prehistory.

**Elevation.** While developing special relativity, Einstein elevated Lorentz invariance to a principle. Earlier we described the process of elevation as a process of hypothetical prioritization. Friedman (2001) argues that the elevation amounts to ascribing an a priori character to the principle of Lorentz invariance. But wouldn't that be in conflict with the principle being testable by interferometer experiments à la Michelson-Morley? Friedman responds that the Michelson-Morley experiments do not exactly show that the light principle has to hold, because the famous null result can also be accommodated by the Lorentz-Fitzgerald competitor theory which works within a classical spatio-temporal structure. The null-result, according to Lorentz and Fitzgerald, is a mere empirical fact, whereas Einstein uses it "as the basis for a radically new spatio-temporal coordination; for Einstein uses his light principle empirically to define a fundamentally new notion of simultaneity" (88). The question of testability and Friedman's account will be discussed in more detail in Section 5.

**Formalization.** With Einstein's elevating Lorentz invariance to the status of a principle the development did not end. Darrigol (2006, 22) argues that "some features that today's physicists judge essential were added only later." In particular, Hermann Minkowski and Arnold Sommerfeld developed the 4-dimensional notation and the relativistic tensor formulation of electromagnetism only after Einstein had developed the principle. This kind of formalization is particularly strong in the sense of providing a *principle framework*, as argued above.

**Generalization.** The most notable efforts of generalization are Einstein's attempts to apply the principle to non-inertial systems, which lead the way to the development of general relativity.

**Challenge.** Challenges to Lorentz invariance are mostly theoretical and specifically motivated in the context of developing theories of quantum gravity, according to Liberati (2013), because many models involve discretization of spacetime, which poses a challenge to the continuous symmetries of Lorentz invariance. However, such challenges typically refer to violations of Lorentz invariance at the Planck scale and thus may still require significant efforts to be tested. A notable experimental challenge to Lorentz invariance was provided in 2011 by the OPERA (Oscillation Project with Emulsion tRacking Apparatus) experiment which suggested a signal for superluminal propagation of neutrinos. Liberati et al. argue that despite these being false detections, the OPERA affair did play an important role in propelling further activity in Lorentz breaking phenomenology which provided useful insights for future searches" (Liberati 2013, 11).

## 3.2 Mach's principle

Unlike Lorentz invariance, Mach's principle has had a much less straightforward development. Its precise statement and interpretation are still under debate (Barbour and Pfister 1995; Fay 2024), as are its successes. It roughly states that the inertia of a body is determined through interactions with other bodies.

**Prehistory.** Ernst Mach's (1915) positivist approach criticized Newton's conclusions, particularly those supported by Newton's spokesman Samuel Clarke, regarding the substantival nature of space and time. According to Norton (1995), Mach did not regard his critique as being based on a principle but rather as a reasoned argument challenging the absolutist underpinnings of Newton. He never explicitly defined it; it was only indirectly suggested, leaving it ambiguous. The principle was later clarified by his followers, with some even claiming to have conceived it independently. The principle, not yet considered a principle, "was a fringe idea" and "had no foundations because of the failure of every experimental test actually tried." (Norton, 1995; 10). So, Mach's principle does not look back to a prehistory within which it received much conceptual or empirical support.

**Elevation.** Nevertheless, Einstein was heavily influenced by Mach's writings. He stated the concept for the first time explicitly in the form of a principle and made it a cornerstone of his theory of general relativity. While recognizing that it may contradict some solutions of the field equations, he then used the principle in order to rule out certain solutions (Einstein 1918; 243). It is interesting to note, as Norton states, that Mach's principle was "often opposed by those who would become Einstein's most ardent supporters." Moreover, Norton argues that the scientific community was "largely uninterested in the proposal" (Norton, 1995; 10).

**Formalization.** Various physicists aimed at providing more concrete formulations and formalizations of the principle. For example, Barbour and Pfister (1995) in their volume on Mach's principle distinguish 21 different formulations of the principle in their index and devote a whole chapter to the question "What is the Machian program?" Similarly, Bondi and Samuel (1997) count eleven different formulations, some of which are incompatible. Among the various formulations of the principle, there are indeed different types of formalization. However, the precise nature of the formalization is also a matter of debate. For example, while Barbour claims that "[i]n the Newtonian context, we can see exactly and without any doubt when a particular structure of a theory makes it Machian", Goenner does not seem to agree: "I cannot fully agree with Barbour when he says [this]" (Goenner 1995; 448).

**Generalization.** While there have been attempts to generalize Mach's principle to other domains, the proposed approaches have not received much attention among physicists. In attempts to derive the particle rest mass as some kind of "cosmic derivation" (Barbour et al. 1995; 478), Machian approaches have been considered in the context of quantum gravitational domains.

**Challenge.** Mach's principle is difficult to challenge. Barbour and Pfister stress that "[i]t has […] to be admitted that presently there exists no unequivocal direct experimental confirmation of any 'Machian effect.'" (Barbour and Pfister 1995; 346) and King goes so far as to claim that "the problem is that we cannot observe enough of the universe to be able to say whether Mach's principle is correct" (King 1995; 246). So, the lack of empirical support, combined with a difficult prospectus, makes it difficult to assess the extent to which the principle can be challenged at all. We will get back to this point in Section 5.

## 3.3 Naturalness

There are many definitions of the naturalness principle. It has been described as a prohibition against fine-tuned parameters (Barbieri and Giudice 1988) or as being associated with the autonomy of scales (Susskind 1979; Williams 2015). More specifically, this means that the physics at relatively low energy scales does not depend sensitively on the physics at relatively high energy scales. The naturalness principle has gained prominence in the context of the Standard Model Higgs boson, because the boson's character of a scalar particle (a particle with spin 0) is claimed to cause a violation of naturalness. Unlike Lorentz invariance, the current status of naturalness as a principle in high-energy physics is strongly contested.

**Prehistory.** Considerations regarding the potentially problematic character of very small or very large dimensionless parameters go back at least to Dirac (1938). Moreover, the principle is claimed to have played a constructive role even before its explicit elevation to a principle, for example, in the context of the charm quark mass prediction. That light scalar particles could be problematic from a theoretical point of view was recognized well before explicit discussions of a naturalness principle emerged (Wilson (1971); see also the discussion by Williams (2019) and Borrelli and Castellani (2019)).

**Elevation.** In the early 1980s, physicists suggested elevating naturalness as a principle for developing new theories (Susskind 1979; 't Hooft 1980). They observed that a scalar boson would require delicate cancellations of quantum corrections, a violation of naturalness that would be prevented by new physics not predicted by the Standard Model. Among other theories this was taken to be a motivation especially for supersymmetry (Fischer 2024b), while the Standard-Model Higgs boson assumed the character of an ad-hoc element of this theory (Friederich et al. 2014).

**Formalization.** There have been many attempts to formalize the naturalness principle in the sense of a restriction to the amount of fine tuning in a theory that is permissible. One of the first measures of fine tuning is the Barbieri-Giudice measure (1988), that quantifies the sensitivity of the theory's output parameters at the electroweak scale on the theory's input parameters and requires that the sensitivity do not exceed a threshold value of 10. This threshold-value is related to Dirac's Large Number Hypothesis.

**Generalization.** The naturalness principle was originally targeted at the Higgs boson. However, soon it was realized that it is potentially applicable also to other sectors in fundamental physics, such as the strong CP problem or the cosmological constant problem, and retrospectively to the charm quark mass and the neutral kaon system. Some of these generalizations provided fertile ground for challenging the naturalness principle.

**Challenge.** Aside from the challenges arising due to its generalization to other domains discussed above, the lack of a unique formulation and a proper formalization has led to challenges of the naturalness principle ever since its elevation (Borrelli and Castellani 2019; Harlander and Rosaler 2019; Rosaler and Harlander 2019; Hossenfelder 2021). Nevertheless, naturalness has nourished the expectation that non-Standard-Model physics would be detected in experiments currently performed at the Large Hadron Collider (LHC). However, no such new physics was found. On the contrary, the Standard Model received yet another experimental confirmation through the discovery of a scalar particle and the subsequent corroboration that this is indeed the long-sought Higgs boson. This situation represents a severe challenge.

In various instances, such challenges have been responded to by changes to the principle and how it is being employed. For example, the boundary of permissible degrees of fine tuning have been repeatedly shifted to higher values. Wells (2023), for example, permits fine tuning up to $10^6$. Moreover, there has been a shift towards using quantitative measures of fine tuning not to exclude theories but instead compare theories regarding their "fine tuning price" (Grinbaum 2012). Nevertheless, it is fair to say that the discovery of the Higgs boson and the associated elimination of the ad-hoc allegation, combined with the lack of discovery of new physics, constitutes a challenge to the naturalness principle that diminished its role as a driving force for the construction of new physics models (Friederich et al. 2014).

## 3.4 The Perfect Cosmological Principle

In contrast to the previous examples, the Perfect Cosmological Principle (PCP) can be seen as an example of a prominent principle that has nevertheless failed.[5] The PCP states, in one formulation, that the universe is not only homogeneous and isotropic, but also stationary.

**Prehistory.** The Perfect Cosmological Principle has several origins and played a central role in the development of steady state cosmological theories in the late 1940s. The roots of the idea of a static cosmology can be traced back to Einstein's seminal 1917 paper on cosmology, in which he proposed a static universe. A static view was an empirically reasonable assumption at the time, as Hubble's empirical evidence for an expanding universe had not yet been proposed (O'Raifeartaigh et al. 2017). Another fundamental influence on the PCP is the cosmological principle that the universe is isotropic and homogeneous. This principle served as the basis for Bondi and Gold's later formulation of the PCP, which extended the cosmological principle to space *and time.*

**Elevation.** As mentioned above, the PCP was elevated to a principle by Bondi and Gold (1948) by extending the cosmological principle: "[t]his combination of the usual cosmological principle and the stationary postulate we shall call the perfect cosmological principle, and all our arguments will be based on it. The universe is postulated to be homogeneous and stationary in its large-scale appearance as well as in its physical laws." Bondi and Gold not only relied on a kind of empirical reasoning that considered the above properties to be part of the large-scale appearance but also postulated them to be general features of the physical laws (see p. 254). While Hoyle (1984) did not motivate the PCP in detail he made it explicit that he wanted to follow a cosmological principle "in its wide sense" (p. 372), extending the equivalence to different times.

---

[5] We thank an anonymous referee for suggesting that we include an example of a failed principle. Another example, but with a much more complex history, is the ether hypothesis (Cantor and Hodge 1981).

**Formalization**. While Bondi and Gold used the de Sitter metric to develop a stationary model, Hoyle modified the field equation of general relativity. However, nothing in the above formulation of the PCP requires a specific way of realizing the PCP, and so various other approaches have been proposed to formalize the principle in terms of modifications of Einstein's field equations, to matter creation without modification of the field equations (McCrea, 1951), to other mechanisms for a static universe. Thus, the principle itself was not strictly speaking formalized, but played the role of a selection rule of relevant cosmologies that satisfy the properties of PCP.

**Generalization.** The PCP is itself a straightforward generalization of the cosmological principle: Any theory that satisfies the PCP automatically satisfies the cosmological principle. There have not been many attempts to generalize the principle itself because it was confronted with empirical data early on. However, the principle was later generalized in the sense of domain generalization, to see if it could accommodate the observations of the CMB (Hoyle et al. 1993) or eternal inflation (Aguirre and Gratton, 2002).

**Challenge.** The theories based on this principle were confronted early on with empirical data. After some initial agreement between the predictions of Bondi and Gold's model and some empirical data on the density of nebulae (Bondi and Gold, 1948, Sect. 3), this success did not continue. In particular, the observed cosmic microwave background challenged the steady-state models, while at the same time supporting the modern big bang model. This conflict with empirical data led to a slow decline in the confidence of cosmologists in the PCP.

In summary, the above examples are quite different kinds of scientific principles. Applying the life-cycle template puts these differences into clearer focus. In what follows we will characterize the differences in more detail by looking at the life cycle as a diagnostic tool and at the testability of scientific principles.

# 4    The life cycle as diagnostic tool

The first application of the life cycle we consider here is as a diagnostic tool. We argue that under certain circumstances the degree to which a principle has gone through the above-described processes can indicate the principle's reflecting features of nature. The basic idea is that (i) principles that reflect features of nature will count as established principles and that (ii) the degree to which a principle is established will be indicated by the processes of the life cycle. By "established", we mean a broad acceptance[6] of the principle among physicists. For example, there are principles that hardly any serious physicist would doubt reflect features of nature, at least within a certain domain, among them the relativity principle, for example, or the Pauli exclusion principle.

(i) First, our argument assumes a strong connection between the principles that are established in a scientific community and the principles reflecting features of nature. Let us refer to this as the *assumption of confidence* – because it expresses confidence in the reliability of scientific inquiry at the scale of research communities.

The assumption of confidence has two parts. First, we assume that if a scientific principle is established in a scientific community, then the principle is highly likely to reflect features of nature. This expresses the confidence that—on a broad

---

[6] Note that we follow Laudan (1978) in distinguishing between a context of acceptance and a context of pursuit. When we say that a principle is established, we mean that it is broadly accepted. This differs from a principle that is endorsed merely as pursuit-worthy heuristic.

scale—scientists are successful at widely accepting only principles that are reflective of their research objects. While we believe that this is a strong assumption, we also see that there are paradigmatic cases of physical principles that encourage such confidence, as the mentioned examples of relativity and the Pauli principle.

The second part of the assumption of confidence is concerned with the reversal of that conditional: if a principle reflects features of nature, then it will be established among physicists. Now, there is a sense in which this is certainly not true: there are many principles that physicists simply do not know but that may well reflect important features of nature. For example, the relativity principle describes important features of nature, and it did so even at a time when no physicists were aware of this.

Here we restrict our discussion to cases that physicists are aware of. And here the assumption of confidence is that principles that have gained such awareness reflect features of nature only if they are established among physicists. Now, obviously not any principle that has come to the awareness of physicists and that reflects features of nature is immediately recognized by all physicists as such a successful principle. Yet the assumption of confidence expresses the idea that after a certain period of probation a principle will either have to be established or otherwise is probably not reflective of features of nature (how long that period of probation is and what kind of research efforts are involved is, of course, a difficult question that will not be answered on a general level).

(ii) Second, we argue that the life cycle provides us with criteria which allow one to judge the degree to which a certain principle is "established" in the physics community at a certain point in time. The idea is that one can do this without looking too closely at specific theories that instantiate those principles. To some extent, what we are doing resembles the methods used in genetics: by tying certain genetic structures to certain medical conditions, it is possible to make predictions even without knowledge of the causal chain that leads from the microscopic DNA to the macroscopic effects.

Any physical principle starts as a conjecture, but already the prehistory that led to this conjecture can give rise to a varying degree in the acceptance of the principle. Earlier we argued that after its elevation, a principle can undergo a number of further processes. Here we identify "formalization" and "generalization" as strongly indicative for a principle's status within the community. On the other hand, a principle can be challenged to certain degrees, which generally indicates a deterioration of its acceptance within the community.

To illustrate the use of the life cycle as a diagnostic tool, let us revisit the three examples discussed in the foregoing section.

First, **Lorentz invariance** is one of the prime examples for an established principle. As we have seen here, this is indeed supported by the degree to which it has undergone the various processes in the life cycle.

It was based on a compelling prehistory which was puzzled by inconsistencies between Galilean relativity and Maxwell's equations for the propagation of light; with the associated tensor calculus acting as a principle framework, it has reached the highest level of formalization; its generalization to non-inertial frames has led to one of the most successful theories in modern physics; and finally, its challenges are either based on experimental errors, or restricted to domains which are far beyond current testability.

**Mach's principle** cannot look back on a successful prehistory. It was considered a "fringe idea", and even after Einstein elevated it to a principle, it remained highly controversial. The ambiguity of its formulation leads to the observation that even when a formalization existed, it did not succeed in convincing the community to rely on it, even among supporters of the principle. These problems also underlie the difficulty of assessing the generalization program and the ways in which it can be challenged. Here we can see that the failure of the principle to go through the various processes is closely related to the lack of a broad agreement in the scientific community about the status and value of the principle.

Next, consider the **naturalness principle**. The prehistory of the naturalness principle is quite convincing. After all, up until the discovery of the Higgs boson, no candidate for an elementary scalar was known. This is quite remarkable: why should there be no elementary field which transforms under the simplest representation of the Lorentz group? Wilson's (1971) paper seemed to contain a quantitative explanation for this. On the other hand, experimental precision data seemed to suggest the existence of such a particle, and supersymmetry seemed to provide a solution which circumvented Wilson's argument.

However, already the existence of various notions and formulations of the naturalness principle indicates that, up to now, its formalization is still at a very vague level. This may also be the reason for a lack of generalization: it remains restricted to the perturbative regime of quantum field theory. And finally, the challenges it has faced over the years, initially based on theoretical arguments, and eventually from the failure of confirming the associated expectations in experiment, have led to a sharp drop of the naturalness principle's significance in the scientific community.

The **perfect cosmological principle** is an interesting example of a principle that was once very prominent, but had a comparatively short life cycle. It did not rely on a prehistory that provided much empirical justification for relying on it. Although it was explicitly elevated, it lacked clear formalization, playing more the role of a selection rule. Nevertheless, its adherents' efforts to establish an empirically testable cosmology (see Sect. 5.2) led to a series of challenges that led to its rather early demise as a viable principle in cosmology.

A potential worry here is that there are scientific principles that have been employed widely despite their somewhat vague character. As examples one may think of renormalizability, the naturalness principle, or the generalized correspondence principle.[7] However, we would contest that such principles have ever achieved the status of established principles of physics as described above (see fn 5). While physicists have employed them, they have typically endorsed them as guiding principles rather than principles of nature. Flagging a principle as a guiding principle usually highlights their heuristic role in the context of theory pursuit but does not imply that they are accepted principles of nature. For a more detailed discussion of the status of these principles as guiding principles and the distinction between guiding principles and accepted principles of nature see Fischer (2024a).

---

[7] We thank an anonymous referee for raising this point.

In summary, the first potential application we see for the life cycle of scientific principles is as a diagnostic tool. Under the assumption of confidence, the degree to which a principle has gone through the above-described processes indicates the principle's reflecting features of nature. This kind of diagnosis may be helpful for a better understanding of past episodes of reasoning with principles. But it may also well be useful as a heuristic for current and upcoming scientific principles.

# 5    The life cycle and the testability of scientific principles

We have seen in the previous sections that principles can undergo different processes. Furthermore, a principle can undergo these processes very differently. In this section we want to consider the extent to which the life cycle of a principle is related to the question of whether a principle can be tested. To be clear, the issue of testability warrants a paper of its own, especially since it is a disputed topic. The main aim here, however, is to provide a plausible proposal for how one might attempt to think about principle testability and see how elements of the life cycle of the principle may help with answering this question.

A potential worry for the question of testability is that principles are sometimes assumed to be exactly those assumptions in a physical theory that cannot be directly tested. Consider, for example, Friedman's (2001) discussion of the role of empirical evidence in evaluating the relativistic light principle (2001, 86f), which we already mentioned in Sect. 3.1. One might think that the Michelson-Morley experiment can be employed to test the assumption that light has the same invariant velocity in all inertial frames. However, according to Friedman, this is not the case. The experiment can "in no way be viewed as an empirical test or crucial test" of special relativity with respect to its theoretical alternatives because in the "Lorentz-Fitzgerald competitor theory to special relativity the very same empirical fact is incorporated within an essentially classical spatio-temporal structure" (87). We believe that the constancy of the speed of light principle of special relativity is one of the most successful principles. If even this paradigmatic example of a successful principle is not testable, how can testability be taken to be an issue for principles in the first place? So, before we can relate the question of testability to that of the life cycle of a principle (Sect. 5.2), we need to understand what it could mean to test a principle in the first place (Sect. 5.1).

## 5.1 Testing principles

The testability of scientific principles is undoubtedly an important question from the perspective of scientific practice.[8] If we take a prominent account of confirmation in current philosophy of science, Bayesian confirmation theory, a hypothesis is confirmed if the probability of the viability of the hypothesis increases in the light of some evidence, and it is disconfirmed if the probability decreases. The probability is updated based on Bayes' theorem. Since hypotheses (together with additional assumptions) ideally predict the empirical data that are supposed to test the theory, there is a direct probabilistic dependence between the empirical data and the hypothesis, and we can confirm or disconfirm the hypothesis based on the observation.

There are now two issues about the testability of a principle that need to be addressed. First, how do we "confirm" a principle when it does not by itself imply the empirical data? And second, if we have significantly disconfirmed a theory that satisfies the principle, what does this tell us about the viability of the principle when there are many more viable theories that satisfy it? Both are rather complicated issues, which we will drastically simplify for the purposes of this paper, where the main aim is to illustrate the fruitfulness of the life-cycle perspective for these questions.

---

[8] C.f. with the typical attitude in physics, where one assumes that the "final judge of any physical principle [...] is experiment" (Pfister 1995, 364).

The first problem is well known in the confirmation theory debate and is related to the Quine-Duhem thesis. A principle may not by itself imply empirical data, but it does so as part of the theory and other assumptions from which jointly one may predict something. One might now ask how to judge the viability of the principle in isolation in cases of confirmation and disconfirmation. This is a question that has received a clear answer from a Bayesian perspective (see, e.g., Dorling (1979) for the classic treatment). In very simple cases of confirmation, we can consider a distributed confirmation of the individual assumptions and thus of the principle. Similarly, in cases of disconfirmation, we may see a similar kind of distribution of disconfirmation among the assumptions and the theory. Both depend strongly on the priors associated with them. This is an issue we will return to in the next subsection, when we speak of the prehistory and the elevation processes of a principle.

The second issue is more intricate. Let us assume we have disconfirmed a theory in the above sense to the extent that we are willing to give up on this theory. This does not necessarily need to be bad news for the principle, as there might be many more viable theories that satisfy it, which have not been tested yet. In this case, we would not want to say that the principle has already been disconfirmed. So, we see that there is an additional dimension that needs to be considered when we want to test a principle, namely the set of theories that satisfy the principle. This theory-space perspective illustrates the meta-theoretical nature of principles, which we may want to impose on any theory of a certain domain. How can we incorporate this theory-space dimension of principle testing?

In order to assess the viability of a principle, we need to have a handle on the corresponding space of theories that satisfy it. Note that we may only be aware of a single well-confirmed theory that satisfies the principle, in which case the principle itself is considered to be supported. However, we may also have proposed many untested theories that satisfy a particular principle, in which case the disconfirmation of a single theory, while disconfirming the principle, may not be so significant in our assessment of the principle.

The idea that the empirical viability of a hypothesis may depend on the space of viable theories is not new. It relies on an underlying idea prominent in (Dawid 2013) and framed within a Bayesian framework in (Dawid et al. 2015), where it was shown that, under certain conditions, finding no alternatives to a scientific theory that can satisfy some constraints is meta-empirical evidence for that scientific theory. In other words, some evidence can confirm a theory, even though it is not directly implied by the theory (Dardashti and Hartmann 2019). Here we do not rely on the no-alternatives argument. Instead, we merely rely on the Bayesian result of that account that puts constraints on scientific underdetermination. These are constraints on the space of viable theories which can provide support for the empirical adequacy of the hypothesis. In other words, one can assess the empirical adequacy of a hypothesis, in our example a principle, by assessing the corresponding theory space.[9]

Let us make the connection between a principle and its corresponding theory space a little more precise. Since principles are satisfied by theories, we can associate each principle with a corresponding theory space containing the theories that satisfy the principle. The point now is that by testing theories within that principle's theory space, we can empirically assess the principle.

Scientific principles are, however, rarely considered in isolation, when we are concerned with the corresponding theory space. So, rather than exploring the whole theory space compatible with a specific scientific principle, additional assumptions are employed to restrict the corresponding theory space. In scientific practice these could be other principles, theoretical virtues such as simplicity, various specific model assumptions, and known empirical constraints. In cases where we are concerned with the testing of the scientific principles, we of course need to reduce these to the minimal assumptions necessary, something Dawid (2013) calls "scientificality conditions".

---

[9] The link between testability and the corresponding theory space has been highlighted in the context of meta-empirical theory evaluation (Dawid, 2013) but plays an equally important role in the context of empirical evaluation (Dawid 2018). We will not go into further detail here on the concept of theory space, which is still under debate (see, for example (Dardashti 2019)). The claims made here will be independent of the specification of the concept of theory space.

To illustrate the testability of a principle within this theory-space perspective, consider the case where we would have perfect knowledge of the theory space corresponding to a principle (an unattainable goal, of course). If all theories in this space could be empirically disconfirmed, one could argue that the principle in this case is all but excluded. Of course, the usual Quine-Duhemian concerns about falsifiability apply here as well, compounded by the additional uncertainty about the corresponding theory space and the scientificality conditions that constrained it. The theory space of the principle is thus an additional element to be considered in the epistemic evaluation of a principle. One can now imagine various other scenarios to illustrate the relation between testability of the principle and its relation to theory space. Consider, for instance, a highly constraining principle whose corresponding theory space allows for only one theory. In that situation the confirmation or disconfirmation of the theory will be identical to that of the principle. If a principle allows for a huge number of theories, the confirmation or disconfirmation of any individual theory will just have a very small impact on the principle's assessment. We will not further elaborate on these details, as we just aim to make plausible the overall approach and its relation to the life cycle of the principle.

A consequence of this view is that a principle is better testable if the corresponding theory space is better empirically accessible. Obviously, the size of the theory space has a direct effect on the extent to which the principle can then be tested. Of course, it may still be the case that a smaller theory space is actually more difficult to test for reasons that are experimentally contingent. For example, new predictions of the principle's theory space might only be observable at energy scales that are not accessible by current experimental means. But we will not consider these practical aspects of testability.

From now on, when we talk about a principle and are concerned with its testability, two different principles will correspond to two different theory spaces for us. Note that this may be at odds with how it is sometimes treated in scientific practice. This is an issue we will return to in the next section in relation to Mach's principle.

We should also note that even if the described procedure of laying out a theory space may ultimately be used to assess a principle, considerations in scientific practice do not always have this as their primary goal. Instead, they use a principle such as naturalness to motivate certain classes of theories and will abandon the principle if the theories turn out to be disconfirmed. While the end result of the corresponding research processes (accepting or abandoning a particular scientific principle) may be the same in these instances, there can be differences in what the primary epistemic goal is: finding new theories that can be employed to generate specific predictions or finding overarching principles that help generate new theories.

## 5.2 Testability and the life cycle

Let us now see how our concept of testability is related to the life cycle of a principle. We start with the processes of **prehistory** and **elevation**. Both play a crucial role in assessing the future success of a principle. Successful principles usually have a prehistory in which they have received some kind of support that justifies their elevation to the status of a principle. The relativity principle would probably not have been elevated to a principle of special relativity if the Michelson-Morley experiment had turned out differently. It has also been argued that the elevation of naturalness to a principle is supported by successful applications of the principle in the past, a claim that has received some criticism though (Sahuqillo-Carretero, 2019). Thus, if a principle is part of an empirically well-supported theory in its prehistory, i.e. when it has not yet been elevated to the status of a principle, this provides support for an individual theory in the principle's theory space and thus indirect evidence for the principle itself. This prehistorical justification is, of course, carried forward into the other processes. If the principle, once elevated, were not to produce empirically successful theories, there would be disconfirmed theories in the principle's theory space and we would slowly expect our trust to shrink with respect to that principle.

While these aspects are relevant to the validity of the principle, the relationship to the testability of a principle will strongly depend on the specific prehistory. The elevation process explicitly prioritizes the principle from the set of assumptions from which we might wish to generate predictions. This in turn, and depending on the degree of prioritization, provides a degree of immunization against possible disconfirming evidence. This is again an issue taken up by Dorling (1979), who showed that the assumptions within a Lakatosian hard core are less affected by disconfirming evidence than the elements of the protective belt. The extreme case, although its scientific validity is debatable, would be to assign a prior of 1 to the principle, in which case no amount of disconfirming evidence would affect the empirical assessment of the principle. This would model Friedman's attitude to principles. So, the testability of a principle depends strongly on the kind of prioritization associated with the elevation of the principle and that in turn depends on the prehistory, where the principle might have received its justification. A metaphysically motivated principle may receive a higher prioritization compared to a heuristic principle.

Let us now turn to the process of **formalization** and its relation to testability. As we have seen, the process of formalization may occur in various ways. If the satisfaction of a principle is precisely defined and allows for a straightforward procedure to check it, the corresponding theory space is similarly precisely defined. That is, one may more easily judge whether a theory belongs to a principle's theory space. If a principle's formalization is vaguely defined, then the boundaries of the theory space associated with the principle become similarly vague. It will be more difficult to judge whether a theory still satisfies the principle or not. This in turn obscures the epistemic access to the principle, as it becomes unclear what exactly needs to be disconfirmed in order to empirically disconfirm the principle. This leaves some leeway for enthusiasts of the principle to further commit and vary the boundaries of the principle's theory space in order to 'save' the principle from empirical disconfirmation.

As already mentioned, Mach's principle has several competing formulations. Let us assume that there is a fairly precise formalization of one such formulation. This would imply that one can place precise bounds on the theory space corresponding to that formulation. This would increase the testability of that formulation of Mach's principle. However, since this is only one formulation of Mach's principle, the status of Mach's principle *simpliciter* remains unclear because a different formulation may correspond to a different theory space. To illustrate this point, consider (Rindler 1994, 238), who argues forcefully that "the Lense-Thirring effect already seems to imply: one cannot trust Mach!". It is in this context, where Bondi and Samuel (1997) present the various formulations of the Mach principle, that shows that Rindler's argument is indeed an argument against the "Mach10" principle, while the "Mach3" principle is fully consistent with the predictions according to the Lense-Thirring effect. Simply put, as long as there is no unique *formulation* of the principle, there cannot be a unique formalization. And as long as there is no unique formalization, it is very difficult to test *the* principle.

Compare the situation with the formalization of Lorentz invariance. This formalization is particularly strong in the sense of providing a principle framework, as argued above. This framework makes it possible to determine unambiguously whether a theory satisfies Lorentz invariance or not. Thus, the corresponding theory space is precisely defined.

The **generalization** of a principle usually amounts to a delimitation of the corresponding theory space. That is, the extension of the principle to other domains implies a restriction of the corresponding principle's theory space, since theories violating the principle in the extended domain are not considered. Note that this only holds if we assume the theory to cover both domains. This restriction of the theory space goes hand in hand with an increase in the predictive success and thus the empirical assessment of the theory (cf. Myrvold's (2003, 2017) analysis in the context of unification). Thus, a more domain- or principle-generalized principle is usually better empirically testable because the corresponding theory space is more restricted.

For example, Lorentz invariance has been domain-generalized several times. It is now an integral part of virtually any theory of fundamental physics. Each further domain extension of the principle has provided empirical support, thus strengthening the commitment to the principle. On the other hand, it can be argued that while there have been several attempts to domain-generalize Mach's principle, none of them have led to empirically successful theories.

Finally, consider the relationship between **challenges** to a principle and the testability of the principle. First, consider empirical challenges. Obviously, for an empirical challenge to occur, one must assume some degree of testability. Thus, the processes discussed so far set the context in which an empirical challenge can occur. Let us compare two examples to illustrate how our proposed framework characterizes different ways in which an empirical challenge can occur. In 2011, the OPERA experiment announced an empirical challenge to relativity. They claimed to have observed faster-than-light neutrinos (The OPERA collaboration 2012). As the theory space associated with the principle is highly constrained, the impact of the observation could be quite strong. However, it was also the case that the scientific community associated a high prior with the principle. So, instead of abandoning the principle completely, the scientific community followed two paths: expand the theory space by revising other theoretical assumptions or our understanding of Lorentz transformations (Amelino-Camelia 2002; Magueijo and Smolin 2002), or turn the evidence around and doubt the empirical claim made by the experiment. In fact, the supposed discovery turned out to be an error in the experimental set-up (Rubbens 2014), and the continued adherence to the principle turned out to be justified.

The Perfect Cosmological Principle is another interesting case in terms of the relation between challenges and testability. One reason for this was that Bondi, in particular, was very much concerned with the empirical elements of cosmological models, having been influenced by Karl Popper. As Kragh puts it, Popper himself considered the Bondi-Gold-Hoyle theory to be "'a very fine and promising theory,' not because it was true but because it was testable and had in fact been falsified" (Kragh 2013). Thus, most of the proposed models of the PCP made very specific predictions that contradicted the empirical data and were therefore not pursued further (see Kragh (1999, chap. 5.1)). However, since the principle did not unambiguously help to determine the space of models that satisfy it (due to its unclear overarching formalization), it left room for other models that could have been—and were—explored much later (Hoyle et al. 1993). In other words, without the more empirically successful Big Bang model, the challenges of the PCP would probably have affected the life cycle of the principle much more slowly.

Compare this with the current status of the naturalness principle, where one could argue that the LHC results pose a strong empirical challenge to it. In contrast to the above case, the corresponding theory space was not initially highly constrained. This explains why there is not one specific observational result that provides an empirical challenge to the naturalness principle, but that it was the accumulation of decades of empirical data (including the results of previous collider experiments) that continued to disconfirm natural theories within the theory space of the principle. Thus, the LHC data have cumulatively provided an empirical challenge to the principle. However, given the vagueness of the principle discussed above, and the associated imprecision of the boundaries of the principle's theory space, there will always be room for continued adherence to the principle by its proponents.

Let us now turn to the theoretical challenges. They operate at the level of the probabilistic priors we associate with the principle. The discovery of a theoretical challenge to a principle will influence whether the scientist wishes to continue to prioritize the principle to the extent that it was prioritized when it was initially elevated. Once the probabilistic prior of a principle is lowered within the set of assumptions from which we derive predictions, it is more susceptible to empirical challenges of the kind discussed above. More specifically, had there been strong accepted theoretical challenges to the naturalness principle, the scientific community might have judged the data from collider physics to be an empirical challenge to the principle much earlier.

As mentioned above, it may not be the goal of a scientific investigation to test a scientific principle. There may be strong metaphysical or other reasons for relying on a principle independently of any empirical evidence for or against it. In our model, this would simply correspond to effectively assigning a probability of 1 to the principle, immunizing it against possible challenges.

This concludes our discussion on how the life cycle allows us to analyze the different issues involved in the testability of a principle.

# 6    Conclusion

In this paper we have introduced the life cycle of scientific principles as a heuristic for analyzing scientific principles in physics. We have identified prehistory, elevation, formalization, generalization, and challenge as characteristic processes of the life cycle.

Specifically, we have looked at four examples of principles from foundational physics: Lorentz invariance, Mach's principle, naturalness, and the perfect cosmological principle. We have seen that the different epistemic statuses of these principles is reflected by their life cycle. More precisely, the differences are reflected, for example, by the kinds of support these principles have achieved in their prehistory and the degree of formalization that they have achieved.

This suggests that the life cycle may be usefully employed as a diagnostic tool. Going through any one of the included processes is not a necessary or sufficient criterion for a principle's success. Yet, one may still make tentative predictions about the success of a principle based on its performance with regard to the processes which are part of our life-cycle template. Moreover, we have argued that the life cycle sheds new light on the testability of scientific principles. If we define a principle extensionally over the set of theories that satisfy it, then the processes indicate how testable a principle is.

# References

Amelino-Camelia, Giovanni. 2001. "Testable Scenario for Relativity with Minimum Length." *Physics Letters B* 510 (1): 255–63. https://doi.org/10.1016/S0370-2693(01)00506-8.

———. 2002. "Relativity in Space-Times with Short Distance Structure Governed by an Observer Independent (Planckian) Length Scale." *International Journal of Modern Physics D* 11 (01): 35–59. https://doi.org/10.1142/S0218271802001330.

Bain, Jonathan. 2016. *CPT Invariance and the Spin-Statistics Connection*. Oxford University Press.

Barbieri, Riccardo, and Gian F. Giudice. 1988. "Upper Bounds on Supersymmetric Particle Masses." *Nuclear Physics B* 306 (1): 63–76. https://doi.org/10.1016/0550-3213(88)90171-X.

Barbour, Julian B., and Herbert Pfister. 1995. *Mach's Principle: From Newton's Bucket to Quantum Gravity*. Vol. 6. Springer Science & Business Media.

Berghofer, Philipp, Jordan François, Simon Friederich, Henrique Gomes, Guy Hetzroni, Axel Maas, and René Sondenheimer. 2023. "Gauge Symmetries, Symmetry Breaking, and Gauge-Invariant Approaches." *Elements in the Foundations of Contemporary Physics*, July. https://doi.org/10.1017/9781009197236.

Bokulich, Alisa. 2008. *Reexamining the Quantum-Classical Relation: Beyond Reductionism and Pluralism*. Cambridge: Cambridge University Press. https://doi.org/10.1017/CBO9780511751813.

Bokulich, Alisa, and Peter Bokulich. 2020. "Bohr's Correspondence Principle." In *The Stanford Encyclopedia of Philosophy*, edited by Edward N. Zalta, Fall 2020. Metaphysics Research Lab, Stanford University. https://plato.stanford.edu/archives/fall2020/entries/bohr-correspondence/.

Bondi, Hermann, and Thomas Gold. 1948. "The Steady-State Theory of the Expanding Universe." *Monthly Notices of the Royal Astronomical Society* 108 (3): 252–70.

Bondi, Hermann, and Joseph Samuel. 1997. "The Lense-Thirring Effect and Mach's Principle." *Physics Letters A* 228 (3): 121–26.

Borrelli, Arianna, and Elena Castellani. 2019. "The Practice of Naturalness: A Historical-Philosophical Perspective." *Foundations of Physics* 49 (9): 860–78. https://doi.org/10.1007/s10701-019-00287-7.

Brown, Harvey R. 2005. *Physical Relativity: Space-Time Structure from a Dynamical Perspective*. Oxford, New York: Oxford University Press. https://doi.org/10.1093/0199275831.001.0001.

Cantor, G. N., and M. J. S. Hodge. 1981. *Conceptions of Ether. Studies in the History of Ether Theories, 1740-1900*. Cambridge University Press. http://archive.org/details/conceptionsofeth0000unse.

Cao, Tian Yu. 2019. *Conceptual Developments of 20th Century Field Theories*. 2nd

ed. Cambrdige, UK: Cambridge University Press.

Carretero Sahuquillo, Miguel Ángel. 2019. "The Charm Quark as *a* Naturalness Success." *Studies in History and Philosophy of Science Part B: Studies in History and Philosophy of Modern Physics* 68 (November):51–61. https://doi.org/10.1016/j.shpsb.2019.06.003.

Crowther, Karen. 2021. "Defining a Crisis: The Roles of Principles in the Search for a Theory of Quantum Gravity." *Synthese* 198 (14): 3489–3516. https://doi.org/10.1007/s11229-018-01970-4.

Dardashti, Radin. 2019. "Physics without Experiments?" In *Why Trust a Theory?: Epistemology of Fundamental Physics*, edited by Karim Thébault, Radin Dardashti, and Richard Dawid, 154–72. Cambridge: Cambridge University Press. https://doi.org/10.1017/9781108671224.012.

Dardashti, Radin, and Stephan Hartmann. 2019. "Assessing Scientific Theories: The Bayesian Approach." In *Why Trust a Theory?: Epistemology of Fundamental Physics*, edited by Karim Thébault, Radin Dardashti, and Richard Dawid, 67–83. Cambridge: Cambridge University Press. https://doi.org/10.1017/9781108671224.006.

Darrigol, Olivier. 2006. "The Genesis of the Theory of Relativity." In *Einstein, 1905–2005: Poincaré Seminar 2005*, edited by Thibault Damour, Olivier Darrigol, Bertrand Duplantier, and Vincent Rivasseau, 1–31. Basel: Birkhäuser. https://doi.org/10.1007/3-7643-7436-5_1.

———. 2021. *Relativity Principles and Theories from Galileo to Einstein*. Oxford, New York: Oxford University Press.

Dawid, Richard. 2013. *String Theory and the Scientific Method*. Cambridge University Press.

———. 2018. "Delimiting the Unconceived." *Foundations of Physics* 48 (5): 492–506. https://doi.org/10.1007/s10701-017-0132-1.

Dawid, Richard, Stephan Hartmann, and Jan Sprenger. 2015. "The No Alternatives Argument." *The British Journal for the Philosophy of Science* 66 (1): 213–34. https://doi.org/10.1093/bjps/axt045.

Dirac, Paul Adrien Maurice. 1938. "A New Basis for Cosmology." *Proceedings of the Royal Society of London. Series A. Mathematical and Physical Sciences* 165 (921): 199–208. https://doi.org/10.1098/rspa.1938.0053.

Dorling, Jon. 1979. "Bayesian Personalism, the Methodology of Scientific Research Programmes, and Duhem's Problem." *Studies in History and Philosophy of Science Part A* 10 (3): 177–87. https://doi.org/10.1016/0039-3681(79)90006-2.

Einstein, A. 1905. "Zur Elektrodynamik bewegter Körper." *Annalen der Physik* 322 (10): 891–921. https://doi.org/10.1002/andp.19053221004.

———. 1918. "Prinzipielles zur allgemeinen Relativitätstheorie." *Annalen der Physik* 360 (4): 241–44. https://doi.org/10.1002/andp.19183600402.

Fay, Jonathan. 2024. "Mach's Principle and Mach's Hypotheses." *Studies in History and Philosophy of Science* 103:58–68.

Fischer, Enno. 2023. "Naturalness and the Forward-Looking Justification of Scientific Principles." *Philosophy of Science*, February, 1–10.

https://doi.org/10.1017/psa.2023.5.

———. 2024a. "Guiding Principles in Physics." *European Journal for Philosophy of Science* 14 (4): 65. https://doi.org/10.1007/s13194-024-00625-1.

———. 2024b. "The Promise of Supersymmetry." *Synthese* 203 (1): 6. https://doi.org/10.1007/s11229-023-04447-1.

Friederich, Simon, Robert Harlander, and Koray Karaca. 2014. "Philosophical Perspectives on Ad Hoc Hypotheses and the Higgs Mechanism." *Synthese* 191 (16): 3897–3917. https://doi.org/10.1007/s11229-014-0504-4.

Friedman, Michael. 2001. *Dynamics of Reason*. Kant Lecture Series. Center for the Study of Language and Information.

Galilei, Galileo. 1632. *Dialogue Concerning the Two Chief World Systems*. Berkeley: University of California Press. [1914]

Goenner, H. F. M. 1995. "Mach's Principle and Theories of Gravitation." In *Mach's Principle: From Newton's Bucket to Quantum Gravity*, 442–57.

Grinbaum, Alexei. 2012. "Which Fine-Tuning Arguments Are Fine?" *Foundations of Physics* 42 (5): 615–31. https://doi.org/10.1007/s10701-012-9629-9.

Harlander, Robert, and Joshua Rosaler. 2019. "Higgs Naturalness and Renormalized Parameters." *Foundations of Physics* 49 (9): 879–97. https://doi.org/10.1007/s10701-019-00296-6.

´t Hooft, Gerard. 1980. "Naturalness, Chiral Symmetry, and Spontaneous Chiral Symmetry Breaking." In *Recent Developments in Gauge Theories*, edited by G.´t Hooft, C. Itzykson, A. Jaffe, H. Lehmann, P. K. Mitter, I. M. Singer, and R. Stora, 135–57. NATO Advanced Study Institutes Series. Boston, MA: Springer US. https://doi.org/10.1007/978-1-4684-7571-5_9.

Hossenfelder, Sabine. 2021. "Screams for Explanation: Finetuning and Naturalness in the Foundations of Physics." *Synthese* 198 (16): 3727–45. https://doi.org/10.1007/s11229-019-02377-5.

Hoyle, Fred, Geoffrey Burbidge, and Jayant V. Narlikar. 1993. "A Quasi--Steady State Cosmological Model with Creation of Matter." *The Astrophysical Journal* 410 (June):437. https://doi.org/10.1086/172761.

Hoyle, Fred. 1948. "A New Model for the Expanding Universe." *Monthly Notices of the Royal Astronomical Society, Vol. 108, p. 372* 108:372.

Jackson, J. D., and L. B. Okun. 2001. "Historical Roots of Gauge Invariance." *Reviews of Modern Physics* 73 (3): 663–80. https://doi.org/10.1103/RevModPhys.73.663.

Janssen, Michel. 2002. "COI Stories: Explanation and Evidence in the History of Science." *Perspectives on Science* 10 (4): 457–522.

King, D. H. 1995. "A Closed Universe Cannot Rotate." In *Mach's Principle: From Newton's Bucket to Quantum Gravity*, 237–48.

Kragh, Helge. 1999. *Cosmology and Controversy. The Historical Development of Two Theories of the Universe*. Princeton University Press. https://press.princeton.edu/books/paperback/9780691005461/cosmology-and-controversy.

———. 2013. "'The Most Philosophically Important of All the Sciences': Karl Popper and Physical Cosmology." *Perspectives on Science* 21 (3): 325–57.

———. 2014. "Historical Aspects of Post-1850 Cosmology." In *AIP Conference Proceedings*, 1632:3–26. American Institute of Physics. https://pubs.aip.org/aip/acp/article-abstract/1632/1/3/836053.

Laudan, Larry. 1978. *Progress and Its Problems: Towards a Theory of Scientific Growth*. Berkely: University of California Press.

Liberati, S. 2013. "Tests of Lorentz Invariance: A 2013 Update." *Classical and Quantum Gravity* 30 (13): 133001. https://doi.org/10.1088/0264-9381/30/13/133001.

Mach, Ernst. 1915. *The Science of Mechanics: A Critical and Historical Account of Its Development, by Ernst Mach: Supplement to the 3rd English Ed*. Open Court Publishing Company.

Magueijo, João, and Lee Smolin. 2002. "Lorentz Invariance with an Invariant Energy Scale." *Physical Review Letters* 88 (19): 190403. https://doi.org/10.1103/PhysRevLett.88.190403.

Massimi, Michela. 2005. *Pauli's Exclusion Principle: The Origin and Validation of a Scientific Principle*. Cambridge: Cambridge University Press. https://doi.org/10.1017/CBO9780511535352.

Mattingly, David. 2005. "Modern Tests of Lorentz Invariance." *Living Reviews in Relativity* 8 (1): 5. https://doi.org/10.12942/lrr-2005-5.

McCrea, William Hunter. 1951. "Relativity theory and the creation of matter." *Proceedings of the Royal Society of London Series A.* 206(1087): 562-75. https://doi.org/10.1098/rspa.1951.0089

Myrvold, Wayne C. 2003. "A Bayesian Account of the Virtue of Unification." *Philosophy of Science* 70 (2): 399–423.

———. 2017. "On the Evidential Import of Unification." *Philosophy of Science* 84 (1): 92–114.

Norton, John D. 1995. "Mach's Principle before Einstein." In *Mach's Principle: From Newton's Bucket to Quantum Gravity*, 9–55.

O'Raifeartaigh, Lochlainn, and Norbert Straumann. 2000. "Gauge Theory: Historical Origins and Some Modern Developments." *Reviews of Modern Physics* 72 (1): 1–23. https://doi.org/10.1103/RevModPhys.72.1.

Pfister, Hubert. 1995. "Experimental Status - Introduction." In *Mach's Principle: From Newton's Bucket to Quantum Gravity*, 364.

Post, H. R. 1971. "Correspondence, Invariance and Heuristics: In Praise of Conservative Induction." *Studies in History and Philosophy of Science Part A* 2 (3): 213–55. https://doi.org/10.1016/0039-3681(71)90042-2.

Radder, Hans. 1991. "Heuristics and the Generalized Correspondence Principle." *The British Journal for the Philosophy of Science* 42 (2): 195–226. https://doi.org/10.1093/bjps/42.2.195.

Rindler, Wolfgang. 1994. "The Lense-Thirring Effect Exposed as Anti-Machian." *Physics Letters A* 187 (3): 236–38.

Rivat, Sébastien. 2021. "Drawing Scales Apart: The Origins of Wilson's Conception of Effective Field Theories." *Studies in History and Philosophy of Science Part A* 90 (December):321–38. https://doi.org/10.1016/j.shpsa.2021.10.013.

Rosaler, Joshua, and Robert Harlander. 2019. "Naturalness, Wilsonian Renormalization, and 'Fundamental Parameters' in Quantum Field Theory." *Studies in History and Philosophy of Science Part B: Studies in History and Philosophy of Modern Physics* 66 (May):118–34. https://doi.org/10.1016/j.shpsb.2018.12.003.

Rubbens, Peter. 2014. "Evaluation of the OPERA Collaboration and the Faster than Light Neutrino Anomaly." https://philsci-archive.pitt.edu/11127/1/scient_expl.pdf.

Susskind, Leonard. 1979. "Dynamics of Spontaneous Symmetry Breaking in the Weinberg-Salam Theory." *Physical Review D* 20 (10): 2619–25. https://doi.org/10.1103/PhysRevD.20.2619.

Tasson, Jay D. 2014. "What Do We Know about Lorentz Invariance?" *Reports on Progress in Physics* 77 (6): 062901. https://doi.org/10.1088/0034-4885/77/6/062901.

The OPERA collaboration. 2012. "Measurement of the Neutrino Velocity with the OPERA Detector in the CNGS Beam." *Journal of High Energy Physics* 2012 (10): 93. https://doi.org/10.1007/JHEP10(2012)093.

Weinberg, Steven. 1996. *The Quantum Theory of Fields: Volume 2: Modern Applications*. Vol. 2. Cambridge: Cambridge University Press. https://doi.org/10.1017/CBO9781139644174.

Wells, James D. 2023. "Evaluation and Utility of Wilsonian Naturalness." *Lecture Notes in Physics*. 1000 (2023): 41-62 https://doi.org/ 10.1007/978-3-031-32469-7_2

Williams, Porter. 2015. "Naturalness, the Autonomy of Scales, and the 125GeV Higgs." *Studies in History and Philosophy of Science Part B: Studies in History and Philosophy of Modern Physics* 51 (August):82–96. https://doi.org/10.1016/j.shpsb.2015.05.003.

———. 2019. "Two Notions of Naturalness." *Foundations of Physics* 49 (9): 1022–50. https://doi.org/10.1007/s10701-018-0229-1.

Wilson, Kenneth G. 1971. "Renormalization Group and Strong Interactions." *Physical Review D* 3 (8): 1818–46. https://doi.org/10.1103/PhysRevD.3.1818.